\documentclass[aps,prl,floatfix,superscriptaddress,reprint]{revtex4-2}
\usepackage{amsmath} 
\usepackage{eurosym}
\usepackage{amsbsy}
\usepackage{latexsym,epsfig}
\usepackage{graphicx}
\usepackage{dcolumn}
\usepackage{subfigure}
\usepackage{comment}
\usepackage{multirow}
\usepackage{color}
\usepackage{bm}
\usepackage{mathrsfs}
\usepackage{amssymb}
\usepackage{amsfonts}
\usepackage{amsmath}
\usepackage{xspace}
\usepackage{soul}
\usepackage{cancel}
\usepackage{float}
\usepackage{epstopdf}
\usepackage{tabularx}
\usepackage{longtable}
\usepackage[colorlinks=true, letterpaper=true, pdfstartview=FitV, linkcolor=red, citecolor=blue, urlcolor=blue]{hyperref}
\usepackage[normalem]{ulem}
\usepackage[version=4]{mhchem}
\newcommand{\be}{\begin{eqnarray}}
	\newcommand{\ee}{\end{eqnarray}}

\usepackage[english]{babel}
\begin{document}
\title{Fractional topology and multi-period re-quantization in open quantum systems}
% \author{Xi Wu}
% %\email{wuxi5949@gmail.com}
% \affiliation{School of Physics, Henan Normal University, Xinxiang 453007, China}
% \author{Xiang Zhang}
% \footnotemark[1]
% \author{Fuxiang Li}
% \affiliation{School of Physics and Electronics, Hunan University, Changsha 410082, China}

\author{Xi Wu}  \thanks{These authors contributed equally to this work.}
\email{wuxi5949@gmail.com}
\affiliation{School of Physics, Henan Normal University, Xinxiang 453007, China}
\author{Xiang Zhang} \thanks{These authors contributed equally to this work.}
\affiliation{School of Physics and Electronics, Hunan University, Changsha 410082, China}
\author{Fuxiang Li}
\email{fuxiangli@hnu.edu.cn}
\affiliation{School of Physics and Electronics, Hunan University, Changsha 410082, China}

\begin{abstract}
   We study fractional topological numbers in open quantum systems described by the Gorin--Kossakowski--Sudarsha--Lindblad master equation. Under symmetry conditions ensuring quantization, we show that single-valued physical states in momentum space give rise to integer winding numbers that remain integer during time evolution. Fractional values arise when this condition is effectively relaxed, such that the topology is evaluated over a restricted sector or exhibits an effective multi-branch structure. In these cases, the winding number is not quantized over the fundamental Brillouin zone and can depend continuously on system parameters, with discontinuities at purity-gap closings. However, when extended over multiple momentum periods, the winding recovers integer quantization. These effects are illustrated in a Su--Schrieffer--Heeger chain with gain and loss and can be probed in long-range hopping photonic lattices with fractional fillings via Bloch state tomography. Our results provide a unified framework for understanding fractional topology in open quantum systems.
\end{abstract}
\maketitle
%\tableofcontents
\footnotetext[1]{These authors contributed equally to this work.}

%\textit{Introduction}.--In recent decades, the study of open quantum systems has become a central topic in theoretical physics\cite{Breuer2002theory,Altland_Simons_2010,Manzano2012aa,Prosen:2008aa,Prosen:2010ab,Buca:2012aa,PhysRevA.86.013606,PhysRevA.89.052133,PhysRevLett.117.137202,Buca:2020aa,PhysRevLett.125.240405,PhysRevLett.126.110404,PhysRevB.105.064302,PhysRevLett.128.033602,PhysRevLett.130.050401,PhysRevLett.132.070402}. Under the Markovian approximation, the dynamics of the density matrix is governed by the Gorini\UTF{2013}Kossakowski\UTF{2013}Sudarshan\UTF{2013}Lindblad (GKSL) master equation\cite{Jamiokowski1972LinearTW,Lindblad1976generators,Gorini1976completely}
%\begin{equation}
%    \frac{d\rho}{dt}=-i[H,\rho]+\sum_{\mu}(2L_{\mu}\rho L_{\mu}^{\dagger}-\{L_{\mu}^{\dagger}L_{\mu},\rho\})\equiv\mathcal{L}\rho,
    \label{lind}
\textit{Introduction}.--In recent decades, experimental advances in cold atoms and photonic lattices have enabled the precise control over disspation and driving, pushing the study of non-equilibrium dynamics in open systems to the forefront of theoretical condensed matter physics\cite{Breuer2002theory,Altland_Simons_2010,Manzano2012aa,Prosen:2008aa,Prosen:2010ab,Buca:2012aa,MULLER20121,PhysRevA.86.013606,PhysRevA.89.052133,PhysRevLett.117.137202,Buca:2020aa,PhysRevLett.125.240405,PhysRevLett.126.110404,PhysRevB.105.064302,PhysRevLett.128.033602,PhysRevLett.130.050401,PhysRevLett.132.070402}. Under the Markovian approximation, the dynamics of the open system is governed by the Gorini--Kossakowski--Sudarshan--Lindblad (GKSL) master equation for the density matrix\cite{Jamiokowski1972LinearTW,Lindblad1976generators,Gorini1976completely}. %For quadratic systems, formal solutions can be obtained via the third quantization framework\cite{Jamiokowski1972LinearTW,Prosen:2008aa,Prosen:2010ab}.
 Recently, increasing attention has been devoted to topological phases in such open systems\cite{PhysRevLett.124.040401,PhysRevB.103.085428,PhysRevB.106.035408,PhysRevLett.127.250402,PhysRevResearch.4.023036,Leefmans:2022aa,PhysRevB.107.165139,PhysRevX.13.031019}. In contrast to  closed systems, where topology is encoded in eigenstates of the Hamiltonian, in open systems it must be defined at the level of the density matrix. 
 
% The GKSL master equation reads
In the GKSL master equation
 \begin{equation}
    \frac{d\rho}{dt}=-i[H,\rho]+\sum_{\mu}\left(2L_{\mu}\rho L_{\mu}^{\dagger}-\{L_{\mu}^{\dagger}L_{\mu},\rho\}\right)\equiv\mathcal{L}\rho,
    \label{lind}
\end{equation}
where $H$ is the Hamiltonian and $L_{\mu}$ describe the coupling to the environment, non-Hermiticty arises in the Liouvillian superoperator $\mathcal{L}$, which fully determines the system’s evolution. Significant progress has been made in understanding the topology of non-Hermitian Hamiltonians\cite{PhysRevLett.116.133903,PhysRevLett.118.045701,PhysRevB.98.245130,PhysRevLett.120.146402,PhysRevLett.121.086803,PhysRevLett.121.136802,PhysRevLett.123.066404,PhysRevX.9.041015,PhysRevLett.123.170401,PhysRevLett.127.196404,Ding:2022aa,Ashida2020aa,ljvt-w6hw,PhysRevLett.134.146602}. Fractional topological invariants have been widely discussed in non-Hermitian systems, where they are often associated with multi-valued eigenstates near exceptional points(EPs). These observations suggest that EPs of Liouvillian superoperator could provide a mechanism for realizing fractional topology in open systems.

However, this EP-based intuition encounters fundamental limitations. In Lindblad dynamics, topological properties are determined by the density matrix \cite{Bardyn:2013aa,PhysRevLett.112.130401,PhysRevB.91.165140,PhysRevX.8.011035,PhysRevB.104.094104} rather than by the eigenstates of the Liouvillian. As a result, the spectral properties of the Liouvillian, including the presence of exceptional points, do not directly control the topology of the steady state. For general states, the density matrices are not constrained by the biorthogonal eigenstructure underlying EP-based constructions. Moreover, in particular, in the non-Hermitian Su--Schrieffer--Heeger (NHSSH) model, the parameter regime that hosts exceptional points also breaks the quantization of the total Berry phase\cite{PhysRevLett.127.090501}. As we shall see in the following text, the reason is the breaking of inversion symmetry. 

These considerations point to a more general mechanism: fractional topology in open systems arises from the multi-valuedness of the physical state in momentum space, rather than from spectral properties of the Liouvillian. More precisely, this requires that the physical state fails to be single-valued over the fundamental Brillouin zone. Guided by this insight, we construct a class of Lindblad models in which such multi-valuedness is introduced in the gain and loss matrices through fractional momentum, while preserving inversion symmetry in the Hamiltonian, dissipation, and initial states. As a result, fractional winding numbers emerge in the steady states. 

Our results are as follows. Under symmetry conditions ensuring quantization, single-valued gain, damping matrices, and initial states enforce an integer winding number throughout evolution. When the gain and loss matrices become multi-valued, the winding number is no longer restricted to integers, but varies continuously with system parameters and exhibits discontinuities at purity-gap closings. Nevertheless, a remnant structure persists: for states with n branches, the winding number over an extended n-period remains strictly integer, establishing multiperiod re-quantization. The proposed setup is experimentally accessible, e.g., in optical superlattices of ultracold fermions, where the Berry phase of steady and transient states can be measured.

\textit{Model}.—Before introducing the specific setup, we briefly summarize the solutions of the GKSL equation for quadratic fermionic systems. The Hamiltonian is given by $H=\sum_{ij}c_i^\dagger h_{ij}c_j$, where $h_{ij}$ defines the single-particle Hamiltonian. We consider single-particle loss and gain processes described by dissipators $L_\mu^l=\sum_i D_{\mu i}^l c_i$ and $L_\mu^g=\sum_i D_{\mu i}^g c_i^\dagger$, with complex coefficients $D_{\mu i}^{l,g}$. In terms of the correlation matrix\cite{PhysRevLett.123.170401,PhysRevB.106.024310} $\Delta_{ij}(t)=\mathrm{Tr}[c_i^\dagger c_j\rho]$, Eq.~(\ref{lind}) reduces to
\begin{equation}
\frac{d\Delta}{dt}=X\Delta+\Delta X^\dagger+2M_g,
\end{equation}
where the damping matrix is
\begin{equation}\label{XhM}
X(k)=ih^T(-k)-M_l^T(-k)-M_g(k),
\end{equation}
with $(M_g)_{ij}=\sum\mu D_{\mu i}^{g*}D_{\mu j}^{g}$ and $(M_l)_{ij}=\sum_\mu D_{\mu i}^{l*}D_{\mu j}^{l}$. %Both $M_g$ and $M_l$ are Hermitian.  
A general solution is given by
\begin{equation}\label{transient}
\Delta(t)= e^{X t} \widetilde{\Delta}(0) e^{X^{\dagger}t} + \Delta_s,
\end{equation} in which the steady-state correlation matrix $\Delta_s=\Delta(\infty)$ is determined by
\begin{equation}
X\Delta_s+\Delta_s X^\dagger+2M_g=0,
\end{equation}
with the formal solution
\begin{equation}
\Delta_s=2 \int_0^{\infty} e^{X t} M_g e^{X^{\dagger}t} d t .
\label{cs}
\end{equation}

We consider a minimal open-system realization of the Su--Schrieffer--Heeger (SSH) model with single-particle gain and loss processes that introduce fractional momentum dependence. In momentum space, the gain and loss dissipators are given by $L^g_1(k)=\sqrt{\gamma_1}(c^{\dagger}_A+c^{\dagger}_B)$, $L^g_2(k)=\sqrt{\gamma_2}(c^{\dagger}_Ae^{-ik/n}+c^{\dagger}_B)$,$L^g_3(k)=\sqrt{\gamma_0}c^{\dagger}_A$, $L^g_4(k)=\sqrt{\gamma_0}c^{\dagger}_B$,  $L^l_1(k)=\sqrt{\gamma_1}(c_A-c^{}_B)$, $L^l_2(k)=\sqrt{\gamma_2}(c^{}_Ae^{-ik/n}-c^{}_B)$, $L^l_3(k)=\sqrt{\gamma_0}c_A$, $L^l_4(k)=\sqrt{\gamma_0}c_B$,  %$L^l_3(k)=\sqrt{\gamma_0}(c^{}_A+i c^{}_B)$.
%\begin{align}
%L^g_1(k) &= \sqrt{\gamma_1}\,(c^{\dagger}_A + c^{\dagger}_B), \\
%L^g_2(k) &= \sqrt{\gamma_2}\,(c^{\dagger}_A e^{-ik/n} + c^{\dagger}_B), \\
%L^g_3(k) &= \sqrt{\gamma_0}\,c^{\dagger}_A, \\
%L^g_4(k) &= \sqrt{\gamma_0}\,c^{\dagger}_B,
%\end{align}
%\begin{align}
%L^l_1(k) &= \sqrt{\gamma_1}\,(c_A - c_B), \\
%L^l_2(k) &= \sqrt{\gamma_2}\,(c_A e^{-ik/n} - c_B), \\
%L^l_3(k) &= \sqrt{\gamma_0}\,(c_A + i c_B).
%\end{align}
For concreteness, we set $n=3$, while the construction is generalized straightforwardly to arbitrary $n$. The corresponding damping matrix $X(k)$ and gain/loss matrices $M_{g/l}(k)$ take the form
\begin{align}
X(k) &= i\left[(t_1 + t_2 \cos k)\sigma_x + t_2 \sin k \, \sigma_y\right] - \Gamma \mathbb{I}, \label{Xk} \\
M_l(k) &= \Gamma\mathbb{I} - \left(\gamma_1 + \gamma_2\cos \frac{k}{3}\right)\sigma_x -  \gamma_2\sin \frac{k}{3}\,\sigma_y, \label{ml} \\
M_g(k) &= \Gamma\mathbb{I} + \left(\gamma_1 + \gamma_2\cos \frac{k}{3}\right)\sigma_x +  \gamma_2\sin \frac{k}{3}\,\sigma_y. \label{mg1}
\end{align}
where $\Gamma = \gamma_0 + \gamma_1 + \gamma_2$ characterizes the relaxation rate of the system. The fractional momentum dependence enters through $k/n$ in $M_{l/g}(k)$, while the damping matrix $X(k)$ retains the conventional $2\pi$ periodicity and does not depend on the {control parameter $\gamma_i$ $(i=0,1,2)$}.

%For usual  local  models and dissipation, fractional momentum may not arise. $M_g(k)$ of Eq. (\ref{mg1}) in coordinate space may involve highly nonlocal dissipation which is not easy to engineer. Here we give a  resolution via simulation: One can consider another model that has a similar structure
%\begin{eqnarray}\label{SSH3k}
%\sigma_y \\
%& M_{l/g}^{\prime}(k)&=\mathbb{I}\mp\cos(k) \sigma_x\mp\sin (k) \sigma_y .
%\end{eqnarray}
%with $K=3 k$ such that in Eq. (\ref{Xk}) Eq. (\ref{ml}) Eq. (\ref{mg1})  $k$ is replaced by $K$.
%\begin{eqnarray}
%& \label{HK} H(K)=\left(t_1+t_3 \cos K\right) \sigma_x+t_3 \sin K \sigma_y \\
%& M_g(K)=\mathbb{I}+\cos \left(\frac{K}{3}\right) \sigma_x+\sin %\left(\frac{K}{3}\right) \sigma_y . \label{MK}
%\end{eqnarray}
%The true momentum is $k \in[0,2 \pi)$ so $K \in[0,6 \pi)$. In order to simulate the same topology, one fills only $1 / 3$ of the lower band s.t. $k \in[0,2 \pi / 3)$, then $K \in[0,2 \pi)$.  In $K$ space, $H(K)$ indeed has the periodicity $2\pi$ and  $ M_g(K)$ has multiplicity three in  $2\pi$.  Model Eq. (\ref{SSH3k}) can be realized by long-range hopping, see Fig. (\ref{moel}). The exact 1/3 filling fraction can be guaranteed in the setup when the initial state, as an eigenstate of $H'(k)$, has a winding number 1. %Then one can find that fractional winding number appears in steady states. Multi-period re-quantization appears naturally as the lower band is completely filled. 

\begin{figure}[htbp]
	\includegraphics[width=1\linewidth]{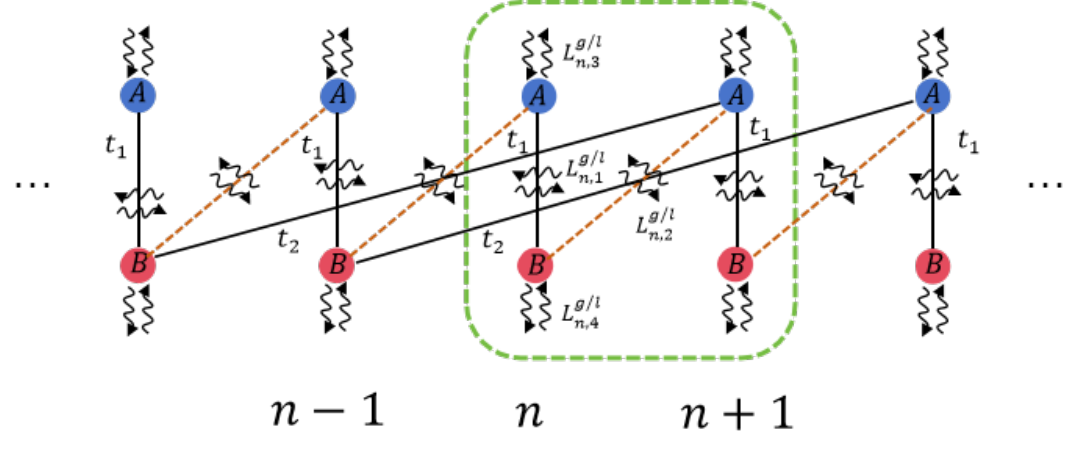}
	\caption{Sketch of the generalized Su-Schrieffer-Heeger model with long-range hopping. Solid lines represent the intra-cell hopping $t_1$ between $(n,A)$ and $(n,B)$ and the inter-cell hopping $t_2$ between site $(n,B)$ and $(n+3,A)$. Black arrows are the fermion loss and gain, described by the dissipators $L_{n,i}^l$ and $L_{n,i}^g$, $i=1,2,3,4$. }	
	\label{moel}	
\end{figure} 
For local models with short-range dissipation, fractional momentum dependence does not naturally arise. In particular, the gain matrix $M_g(k)$ in Eq.~(\ref{mg1}) corresponds to highly nonlocal dissipation in real space, which is difficult to engineer directly. To address this, we introduce a construction that simulates the same topology in a physically accessible way:
{\begin{align}
H'(k) &= \left(t_1 + t_2 \cos(3k)\right)\sigma_x + t_2 \sin(3k)\,\sigma_y, \label{SSH3k} \\
M_{l/g}'(k) &= \Gamma\mathbb{I} \mp (\gamma_1+\gamma_2\cos k)\, \sigma_x \mp \gamma_2\sin k\, \sigma_y.
\end{align}}
Defining $K = 3k$, the above model maps to Eqs.~(\ref{Xk})--(\ref{mg1}) with $k$ replaced by $K$. Since the physical momentum satisfies $k \in [0,2\pi)$, the variable $K$ spans $[0,6\pi)$. To recover the same topological structure, we restrict to a momentum-resolved occupation of the lower band within $k \in [0,2\pi/3)$, corresponding to an effective one-third filling in momentum space, such that $K \in [0,2\pi)$. In this representation, $H'(K)$ is $2\pi$-periodic, while $M_g'(K)$ exhibits three values for the same $K$. The restriction to $K \in [0,2\pi)$ then corresponds to selecting a single branch, while the full topology is recovered only when all branches are taken into account.  This model can be implemented using long-range hopping in real space, as illustrated in Fig.~\ref{moel}. The required effective $1/3$ filling can be restricted by initializing the system in an eigenstate of $H'(k)$ with winding number $1$.

\begin{figure*}[htbp]
	\includegraphics[width=1\linewidth]{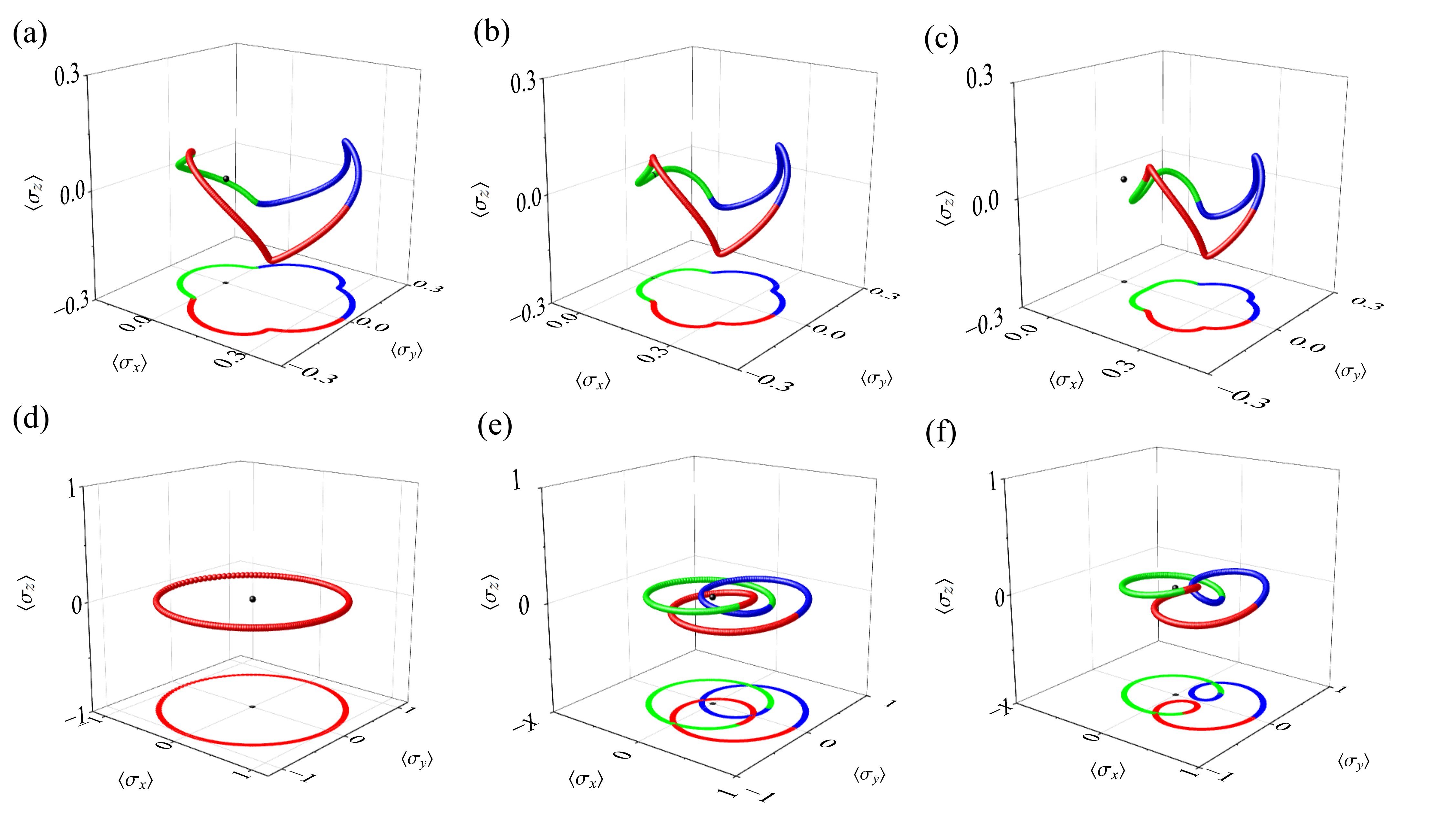}
	\caption{Topological phase transitions of non-equilibrium steady state (a-c) and dynamical phase transitions (d-f) in SSH model for period $6\pi$. Blue, green, and red lines correspond to period $2\pi$, respectively. For (a-c), {the control parameters are $\gamma_1=0.5,1$, and $1.5$}. For (d), different curves overlap with each other and the initial state is topological non-trivial with integer topological invariant $\nu=1$ in 2$\pi$ period. For (e-f), the winding is changed with time and dynamical phase transitions occur after (e) where the topological invariant is changed from integer (e) to fraction (f). For (d-f), the parameter is $\gamma_1=0.5$, ensuring that the steady state is in fractional topology. Other parameters are $t_1=1$, $t_2=2$, $\gamma_0=0.5$ and $\gamma_2=1$.
    }
	\label{ss}	
\end{figure*} 

\textit{General theorem for winding number}.—We first establish a general constraint on the winding number, independent of its specific definition.  If $X(k)$ and $M_{g/l}(k)$ are single-valued in the Brillouin zone, i.e.,
\begin{equation}
X(k+2\pi)=X(k), \quad M_{g/l}(k+2\pi)=M_{g/l}(k),
\end{equation}
then the winding number of the steady state is quantized as an integer. To see this, we expand the steady-state correlation matrix in the Pauli basis,
\begin{equation}
\delta^s_a(k)=\mathrm{Tr}\bigl[\Delta_s(k)\sigma_a\bigr], \quad {a=0,1,2,3}.
\end{equation}
Using Eq.~(\ref{cs}), the coefficients can be written as
\begin{equation}
\delta^s_i(k)=2 \int_0^{\infty} \mathrm{Tr}\left(e^{X(k)t} M_g(k) e^{X^\dagger(k)t} \sigma_i\right) dt\,.
\end{equation}
Since both $X(k)$ and $M_g(k)$ are single-valued, the integrand is periodic in momentum,
\begin{equation}
\mathrm{Tr}\left(e^{X t} M_g e^{X^{\dagger}t} \sigma_i\right)(k+2\pi)
=
\mathrm{Tr}\left(e^{X t} M_g e^{X^{\dagger}t} \sigma_i\right)(k),
\end{equation}
and remains finite for all $t$. It follows that
\begin{equation}
\delta^s_a(k+2\pi)=\delta^s_a(k),
\end{equation}
and therefore the steady-state correlation matrix is itself single-valued,
\begin{equation}
\Delta_s(k+2\pi)=\Delta_s(k).
\end{equation}
Provided that a symmetry-protected geometric phase can be defined (as discussed below), the single-valuedness of the state implies that the winding number is quantized as an integer. For transient states in Eq. (\ref{transient}),  the same conclusion holds provided the initial state ${\Delta}(0)$ is single-valued, since the time evolution preserves this property. Notably, the presence of exceptional points in the non-Hermitian matrix $X$ leads to multi-valued eigenvalu0es and eigenvectors, but does not affect the single-valuedness of $X(k)$ as a function, and therefore does not induce multi-valuedness of the physical state. %This also explains why exceptional-point physics is generally not reflected in steady states\cite{PhysRevLett.134.146602}.

A direct route to fractional winding numbers is to relax the single-valuedness condition by allowing a multi-valued structure in momentum space. Concretely, we consider
\begin{equation}
\begin{split}
    M_{g/l}(k+2\pi)&\neq M_{g/l}(k),\\
    M_{g/l}(k+2n\pi)&=M_{g/l}(k),
\end{split}
\end{equation}
such that the gain and loss matrices are periodic only over an extended Brillouin zone of size $2n\pi$. Here, $n>1$. As a consequence, while the winding number evaluated over the fundamental $2\pi$ Brillouin zone can take fractional values, the winding number defined over the extended $2n\pi$ Brillouin zone remains strictly integer, establishing a multi-period re-quantization of topology.

\textit{Definition of topological invariant}.—To define the winding number for generic mixed states, we introduce a geometric construction of the Berry phase. We map a mixed-state density matrix $\rho$ to a pure-state density matrix $P=|\psi\rangle\langle\psi|$ that preserves the angle coordinates on the Bloch sphere at each momentum $k$, which we refer to as directional purification,
\begin{equation}
\rho(k) \xrightarrow{\text{same angle coordinate}} P=|\psi(k)\rangle \langle \psi(k)| \, .
\end{equation}
The winding number is then defined from the Berry phase of the purified state,
\begin{equation}
\nu=\frac{\Phi}{\pi} = \frac{-i}{\pi}\int_{0}^{2\pi} \langle \psi|\partial_k|\psi\rangle\, dk = \frac{1}{4\pi}\int_S \epsilon_{ijk}\,\hat{\rho}_i \, d\hat{\rho}_j \wedge d\hat{\rho}_k \, ,
\end{equation}
where $S$ is the surface enclosed by the trajectory of $\hat{\rho}_i(k)=\mathrm{Tr}[P \sigma_i]$. Equivalently, $\hat{\rho}_i(k)$ can be expressed directly in terms of the mixed state as $\hat{\rho}_i(k)=\rho_i(k)/\sqrt{\sum_j \rho_j^2(k)}$, with $\rho_i(k)=\mathrm{Tr}[\rho \sigma_i]$. This construction preserves the solid angle and is invariant under different representations (density matrix, correlation matrix, or modular Hamiltonian), providing a natural extension of the geometric phase to mixed states (see Supplementary Material). Moreover, it admits a direct geometric interpretation through the trajectory of $\langle \sigma_x\rangle$, $\langle \sigma_y\rangle$, and $\langle \sigma_z\rangle$ on the Bloch sphere, and corresponds to an observable quantity in real space as the ensemble geometric phase in the thermodynamic limit\cite{PhysRevX.8.011035}.

To ensure quantization of the Berry phase in one dimension, symmetry plays a crucial role. In open quantum systems, chiral symmetry is generally not preserved under time evolution, except in trivial cases\cite{Mao_2024,PhysRevB.110.155141}. To access nontrivial physics in the SSH model, we instead impose inversion symmetry\cite{PhysRevB.108.085126}, originally introduced in the context of unitary dynamics, and extend it here to Lindblad evolution,
\begin{equation}
\sigma_x \mathcal{O}(k) \sigma_x = \mathcal{O}(-k), \quad \mathcal{O}=X, M_{g/l}, \Delta(0,k),
\end{equation}
which is preserved during the evolution (see Supplementary Material for details) and leads to
\begin{equation}\label{sigDeltsig1}
\sigma_x \Delta(t,k) \sigma_x = \Delta(t,-k).
\end{equation}
Expressed in terms of the Bloch vector $\hat{\delta}_i = \delta_i / \sqrt{\sum_j \delta_j^2}$ with $\delta_i=\mathrm{Tr}[\Delta \sigma_i]$ ($i=x,y,z$), this symmetry implies
\begin{align}
\hat{\delta}_x(k) &= \hat{\delta}_x(-k), \\
\hat{\delta}_{y,z}(k) &= -\hat{\delta}_{y,z}(-k).
\end{align}
\begin{figure}[ht]
	\includegraphics[width=0.4\textwidth]{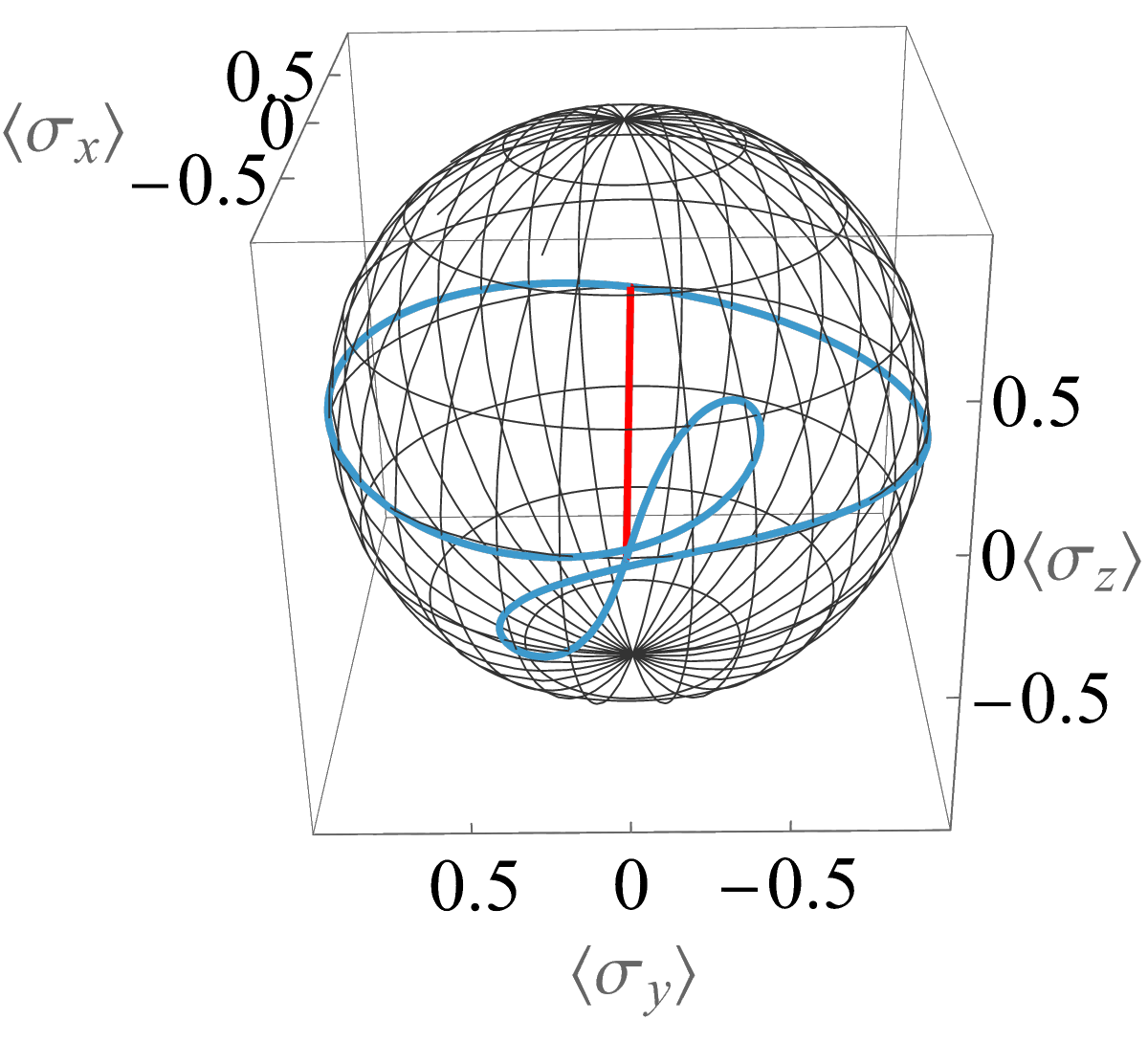}
	\caption{The trajectory of $\hat{\delta}_i(k)$ on the Bloch sphere, symmetric about $k=0$ and $k=3\pi$, divides the surface into two equal parts. The red line is the symmetry axis.}
	\label{windinginversionsym}	
\end{figure} 
Geometrically as seen in Fig. \ref{windinginversionsym}, the trajectory of $\hat{\delta}_i(k)$ on the Bloch sphere is invariant under a $\pi$ rotation about the $x$-axis, resulting in an axis-symmetric contour. The enclosed solid angle is therefore quantized to either $0$ or $2\pi$, implying that the Berry phase takes values $0$ or $\pi$ for a completely filled lower band. However, introducing a term proportional to $\sigma_y$ in Eq.~(\ref{Xk}) breaks this inversion symmetry; in particular, the parameter regime that generates exceptional points in $X$ also destroys this symmetry.

\textit{Topological phase transitions beyond the integer realm}.—We define fractional topology as a winding number that is fractional within a single Brillouin zone but becomes quantized when extended over multiple periods. As a concrete realization, we consider a three-branch structure corresponding to a three-period re-quantization, which can be observed in both steady and transient states. In the presence of multi-valued mappings in momentum space, the winding number is no longer restricted to integers, but depends continuously on the control parameter $\gamma_1$, allowing for exact fractional value $1/3$ under appropriate tuning. We first illustrate this in the steady state. From Eq.~(\ref{cs}), the steady-state correlation matrix $\Delta_s$ is determined by the gain matrix $M_g$, and through Eq.~(\ref{mg1}) the parameter $\gamma_1$ drives a topological transition, as shown in Fig.~\ref{ss} (a-b). {The steady state is topological for $\gamma_1<\gamma_2$ and trivial for $\gamma_1>\gamma_2$,} corresponding to winding numbers $1$ and $0$ over an extended momentum period $[0,6\pi]$, and hence fractional values within the fundamental $2\pi$ Brillouin zone. At the critical point $\gamma_1=\gamma_2$, the trajectory of $(\langle\sigma_x\rangle,\langle\sigma_y\rangle)$ crosses the origin, signaling a topological phase transition. Focusing on a single branch, e.g., the interval $[2\pi,4\pi]$, the trajectory spans an angular range $\Delta\phi=2\pi/3$ in the $\langle\sigma_x\rangle$--$\langle\sigma_y\rangle$ plane; combined with inversion symmetry, this yields a solid angle $2\pi/3$ on the Bloch sphere and a Berry phase $\Phi=\pi/3$, corresponding to an exact fractional winding number $1/3$. 

We now turn to dynamical phase transitions, where the topological invariant evolves nontrivially in time. The main results are shown in Fig.~\ref{ss} (d-f). The projection in the $\langle\sigma_x\rangle$--$\langle\sigma_y\rangle$ plane  display the winding number extracted from $(\langle\sigma_x\rangle(t),\langle\sigma_y\rangle(t))$ over an extended $6\pi$ momentum period, while the three-dimensional plot shows the corresponding three-dimensional winding of $(\langle\sigma_x\rangle(t),\langle\sigma_y\rangle(t),\langle\sigma_z\rangle(t))$. In contrast to previous studies\cite{Mao_2024,PhysRevB.110.155141}, we uncover a dynamical transition from integer to fractional topology. At $t=0$, the initial state is chosen as an eigenstate of the damping matrix $X$ yielding winding number $ 1$ within the $2\pi$ Brillouin zone, as shown in Fig.~\ref{ss} (d), where different momentum branches overlap. As time evolves, the system undergoes a dynamical topological transition [Fig.~\ref{ss}(e)--(f)], after which the winding number becomes fractional when evaluated over a $2\pi$ period. Although the resulting value is not exactly $1/3$, integer quantization is restored when the momentum range is extended to $6\pi$, consistent with the multi-period re-quantization. These transitions are characterized by the closing of the purity gap, where the correlation matrix becomes proportional to the identity at some momentum $k_0$. For steady states, this condition reduces to $M_g(k_0)\propto M_l^T(-k_0)$, while for transient states it corresponds to
\begin{equation}
\frac{d}{dt}\left(e^{X t} \widetilde{\Delta}(0) e^{X^{\dagger}t}\right) = (2-c)M_g(k_0)-cM_l^T(-k_0),
\end{equation}
with $c$ a constant (see Supplementary Material for details).

\textit{Summary and discussion}.--We have proposed a general mechanism for fractional topological number within a single Brillouin zone in open systems that last to steady states, which is multi-valuedness in states.  This fractional number becomes quantized when extended over multiple periods.  The proposed model can be implemented in optical superlattices with engineered long-range hopping using Raman-assisted tunneling \cite{atala2013,PhysRevLett.111.185302,PhysRevLett.111.185301}.%The proposed model can be implemented in a 1D optical superlattice using ultracold fermionic atoms (e.g.\ $^{40}$K or $^{6}$Li) confined in optical potentials.  A tunable SSH-type configuration is realized via staggered tunneling amplitudes $t_1$ and $t_2=0$ controlled by adjusting the relative intensities and phases of the two standing-wave laser fields forming the superlattice \cite{atala2013}.  An additional Raman-assisted tunneling is introduced to couple sites $A(n)$ and $B(n+3)$, thereby generating  $t_3$.  This can be implemented by introducing a momentum transfer $3(2\pi/a)$ through a two-photon Raman process \cite{PhysRevLett.111.185302,PhysRevLett.111.185301}.  
%The amplitude and phase of $t_3$ are controlled by the Raman intensity and detuning, allowing precise tuning of the effective Hamiltonian.

A direct observation can be accessed via standard Bloch state tomography\cite{PhysRevLett.107.235301,PhysRevLett.113.045303,doi:10.1126/science.aad4568,doi:10.1126/science.aad5812}, which has been used to probe topology in quench dynamics\cite{PhysRevB.97.060304,WangCe2017,ChenXin2020,HuHaiping2020,PhysRevB.101.155131,PhysRevB.110.224304,ZHANG20181385,ZhangLong2019,PhysRevA.102.042209,PhysRevA.107.052209,PhysRevA.106.022219}.  One just needs to measure  $ {\rho}_i(t,k)=\text{Tr}\rho(t,k){\sigma}_i$ or $\delta_i(t,k)=\text{Tr}\Delta(t,k){\sigma}_i$, being aware that $\sigma_i$ is not the true spin but a pseudo spin coming from the two-band structure, and plot the trajectory of them in the Bloch sphere for each time $t$ as it evolves. %from the experiment, we measure the Berry phase of directional purified states. This means that for the mixed states depending on momentum $k$ in the Bloch ball, we map them to pure states which have the same solid angles as the corresponding mixed states.  Mathematically, we can define
%\begin{eqnarray}
%   \rho_P(k)= \Delta_P(k)=\frac{\Delta(k)}{|{\rm Tr}[\Delta(k)\boldsymbol{\sigma}]|}\,.
%\end{eqnarray}

\textit{Acknowledgements}.--X Wu thank Prof. Yan He and Prof. Keyu Su for invaluable discussions. This work was supported by the National Natural Science Foundation of China (No. 12275075) and Doctoral scientific research foundation of Henan Normal University(Grants No. 5101029170913)..

\medskip
\appendix

\section{Solid angle is preserved between the correlation function, the modular Hamiltonian and the density matrix}\label{solidangle}
The fact that the solid angle is preserved among the correlation function $\delta$, the modular Hamiltonian $K$, and the density matrix $\rho$ validates the uniqueness of our construction of fractional winding number. Note that the solid angle of the density matrix $\rho$ as an example, is given by
\begin{eqnarray}
   \Omega=\int_S \epsilon_{ijk}~\hat{\rho}_i d\hat{\rho}_j \wedge d\hat{\rho}_k \, .
\end{eqnarray}
In the following we prove a more generic result: let $F=f(\Delta)$ be a matrix function of $\Delta$, then 
\begin{eqnarray}\label{Fideli}
    \hat{F}_i=\hat{\delta}_i, i=1,2,3
\end{eqnarray}
where $ \hat{F}_i=\frac{F_i}{|\vec{F}|} $,  $ F_i=\frac{1}{2}\text{Tr}\left(F \sigma_i\right)$, $  \hat{\delta}_i=\frac{\delta_i}{\left|\vec{\delta}\right|},  \delta_i=\frac{1}{2} \text{Tr}\left(\Delta \sigma_i\right) \,.
$ This result tells us that the matrix function operation does not change the angle coordinate of the $\sigma$-average of a matrix. After integration, this means that the Berry phase is invariant under such operation.

Proof:
Define
\begin{eqnarray}
\bar{\Delta}:=\Delta-\delta_0 \mathbb{I}, 
\end{eqnarray}
then 
\begin{eqnarray}
    \bar{\Delta}^2=\left|\vec{\delta}\right|^2 \mathbb{I}
\end{eqnarray}
For simplicity we denote $f({\Delta})$ by $f(\bar{\Delta})$ 
\begin{eqnarray}
 F=f(\bar{\Delta})=f_{\text {even }}(\bar{\Delta})+f_{\text {odd }}(\bar{\Delta})\,,\\
 \text {where }\left\{\begin{array}{l}
f_{\text {even }}(\bar{\Delta})=\frac{f(\bar{\Delta})+f(-\bar{\Delta})}{2} \\
f_{\text {odd }}(\bar{\Delta})=\frac{f(\bar{\Delta})-f(-\bar{\Delta})}{2}\,.
\end{array}\right.
\end{eqnarray}
We then take a Taylor expansion
\begin{eqnarray}
f(\bar{\Delta})=\sum_{n=0}^{\infty} \frac{\bar{\Delta}^n}{n!} f^{(n)}(0) 
\end{eqnarray}
  and get
\begin{eqnarray}
f_{\text {even }}(\bar{\Delta})&=&\sum_{n=0}^{\infty} \frac{\bar{\Delta}^{2 n}}{(2 n)!!} f^{(2n)}(0)=f_{\text {even }}\left(\left|\vec{\delta}\right| \mathbb{I}\right) \\
f_{\text {odd }}(\bar{\Delta})&=&\sum_{n=0}^{\infty} \frac{\bar{\Delta}^{2 n+1}}{(2 n+1)!!} f^{(2 n+1)}(0)\nonumber
\\
& =&\sum_{n=0}^{\infty} \frac{({\left|\vec{\delta}\right| \mathbb{I}})^{2 n+1}}{(2 n+1)!!} f^{(2 n+1)} \frac{\bar{\Delta}}{\left|\vec{\delta}\right| }\nonumber\\
&=&f_{\text {odd }}\left({\left|\vec{\delta}\right| \mathbb{I}}\right) \frac{\bar{\Delta}}{\left|\vec{\delta}\right|}
\end{eqnarray}
so 
\begin{eqnarray}\label{fDeltafevenfodd}
  f(\bar{\Delta})   =f_{\text {even }}\left({\left|\vec{\delta}\right| \mathbb{I}}\right)+f_{\text {odd }}\left({\left|\vec{\delta}\right| \mathbb{I}}\right) \frac{\bar{\Delta}}{\left|\vec{\delta}\right|} \,.
\end{eqnarray}
Next we take $\sigma$-averages on both side of Eq. (\ref{fDeltafevenfodd}) and get
\begin{eqnarray}
     {F}_i=\frac{f_{\text {odd }}({\left|\vec{\delta}\right| \mathbb{I}})}{|\vec{\delta}|} \delta_i \,.
\end{eqnarray}
Using the definition of $\hat{F}_i$ and $\hat{\delta}_i$ we finally arrive at Eq. (\ref{Fideli}).

\section{Inversion symmetry during time evolution}\label{Inversym}
In this section, we show that inversion symmetry protect the quantized value  $\pi$ of Berry phase in Lindblad open systems, generalizing the result in quench dynamics in \cite{PhysRevB.108.085126}. Inversion symmetry can be expressed for operators as:
\begin{eqnarray}
    \label{sigXsig}\sigma_x X(k) \sigma_x&= &X(-k)\,\\
     \label{sigMgsig}\sigma_x M_g(k) \sigma_x&= &M_g(-k)\\
      \label{sigDel0sig} \sigma_x\Delta(0,k) \sigma_x&= &\Delta(0,-k) \,
\end{eqnarray}
where $\Delta(0,k)$ is the single-particle correlation in the initial state.

The proof can be done in two steps: first, %we show the equivalence of  inversion symmetry in the correlation matrix and that in the modular Hamiltonian; second, 
we prove that the inversion symmetry is preserved in  the correlation function during the time evolution and in the steady states; second, we show that based on the inversion symmetry, the Berry phase is quantized.
 \if0
The relation between  the correlation function $\Delta$ and the modular Hamiltonian $K$ is
\begin{eqnarray}
     \Delta = \frac{1}{e^{K^T}+1}\Leftrightarrow  K^T=\ln{(\Delta^{-1}-1)}\,.
\end{eqnarray}
Then  the following expressions
\begin{eqnarray}
\sigma_x\Delta\sigma_x &= &\frac{1}{e^{\sigma_xK^T\sigma_x}+1}\,,\\
    \sigma_x K^T\sigma_x&=&\ln{((\sigma_x\Delta\sigma_x)^{-1}-1)}\,%=\ln{(\Delta^{-1}(-k)-1)}=\sigma_x K^T(-k) \sigma_x\,.
\end{eqnarray}
 validate the equivalence of  inversion symmetry in two matrices
\begin{eqnarray}
\sigma_x\Delta(k) \sigma_x =\Delta(-k) \Leftrightarrow \sigma_xK(k) \sigma_x =K(-k)\,.
\end{eqnarray}
\fi
First, we prove 
\begin{eqnarray}\label{sigDeltsig}
    \sigma_x  \Delta(t,k) \sigma_x= \Delta(t,-k)\,.
\end{eqnarray}
From Eq.(\ref{sigXsig}) Eq. (\ref{sigMgsig}) ,
\begin{eqnarray}
    \sigma_x \Delta_s(k) \sigma_x&=&2\sigma_x  \int_0^{\infty} e^{X(k) t} M_g(k) e^{X^{\dagger}(k)t} d t \sigma_x  \,,\nonumber\\
    &=& 2\int_0^{\infty} e^{X(-k) t} M_g(-k) e^{X^{\dagger}(-k)t} d t =\Delta_s(-k)  \,.
\end{eqnarray}
From Eq. (\ref{sigDel0sig})  and
\begin{eqnarray}
     \Delta(0,k)= \widetilde{\Delta}(0,k)+\Delta_s(k)\,,
\end{eqnarray}
we have
\begin{eqnarray}
    \sigma_x  \widetilde{\Delta}(0,k) \sigma_x= \widetilde{\Delta}(0,-k)\,.
\end{eqnarray}
Therefore,
\begin{eqnarray}
    \sigma_x  \Delta(t,k) \sigma_x&=&  \sigma_x  e^{X(k) t} \widetilde{\Delta}(0,k) e^{X^{\dagger}(k)t} \sigma_x  +\sigma_x \Delta_s(k) \sigma_x\nonumber\\
    &=&e^{X(-k) t} \widetilde{\Delta}(0,-k) e^{X^{\dagger}(-k)t}   + \Delta_s(-k)\nonumber\\
    &=& \Delta(t,-k)\,.
\end{eqnarray}

Next, we show the quantization of Berry phase. Eq. (\ref{sigDeltsig})  can be expressed by $\sigma$ averages  $ \delta_a=\text{Tr}(\Delta \sigma_a)\,$ as
\begin{eqnarray}
    \label{}\delta_{0,1} (k) &= &\delta_{0,1} (-k)\,\\
     \label{}\delta_{2,3} (k) &= &-\delta_{2,3} (-k) \,.
\end{eqnarray}
Then for $\hat{\delta}_i=\delta_i/{\delta_j^2}$ where $i=x,y,z$ we have 
\begin{eqnarray}
    \label{}\hat{\delta}_{x} (k) &= &\hat{\delta}_{x} (-k)\,\\
     \label{}\hat{\delta}_{y,z} (k) &= &-\hat{\delta}_{y,z} (-k) \,.
\end{eqnarray}
\begin{figure}[ht]
	\includegraphics[width=1\linewidth]{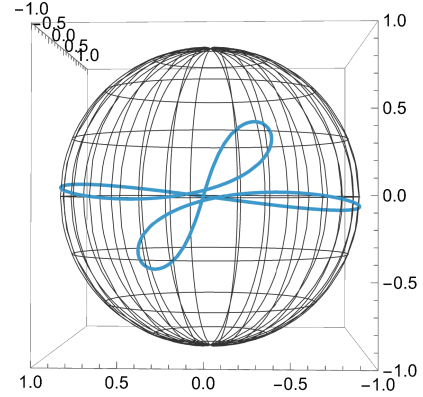}
	\caption{The trajectory of $\hat{\delta}_i(k)$ on the Bloch sphere, symmetric about $k=0$ and $k=\text{half period}$, divides the surface into two equal parts.}
	\label{windinginversionsym1}	
\end{figure} 
The Berry phase is half the area whose contour is given by the trajectory of  $\hat{\delta}_i(k)$ in Fig. \ref{windinginversionsym1}.  The inversion symmetry is translated into a central symmetry in the sphere, that is, a $\pi$ rotation around $x$-axis which makes this contour unchanged. Therefore, the area above this contour is the same as the area below this contour, making them both $2\pi$. So the Berry phase is quantized as $\pi$.

\section{Topological phase transition point}\label{Phasetran}
In this section, we derive the purity-gap closing condition in steady state phase transition and dynamical phase transitions.
Topological phase transition for steady states occurs at\\
\begin{eqnarray}
M _g(k) \propto X(k)+X^{\dagger}(k)=M _g(k)+M_ l^T(-k) \\
\Leftrightarrow M_g{(k)} \propto M_l^T(-k)
\end{eqnarray}
This can be checked as following
\begin{eqnarray}
\Delta_s(k)&=&2 \int_0^{\infty} e^{X t} M_g e^{X^{\dagger}t} d t\nonumber\\
& \propto&\int_0^{\infty} e^{X t}\left(X(k)+X^{\dagger}(k)\right) e^{X^{\dagger} t} d t \nonumber\\
&= & \int_0^{\infty}\left(e^{X t} d(X t) e^{X^{\dagger} t}+e^{X t}  e^{X^{\dagger} t} d\left(X^{\dagger} t\right)\right)\nonumber \\
&= & \int_0^{\infty} d e^{X t} \cdot e^{X^{\dagger} t}+e^{X t} d e^{X^{\dagger}t}\nonumber \\
&= & \int_0^{\infty} d\left(e^{X t} \cdot e^{X^{\dagger} t}\right)=\left.\left(e^{X t} e^{X^{\dagger} t}\right)\right|^{t=\infty}_{t=0}=-\mathbb{I}(k)\,, %\\
%& X=i h+M
\end{eqnarray}
in which the last step is closing of purity gap, meaning the steady state becomes  completely mixed at momentum $k$.

Dynamical topological phase transition occurs at 
\begin{eqnarray}\label{dnpt}
    \frac{d}{d t}\left(e^{X(k) t} \widetilde{\Delta}(0,k) e^{X^{\dagger}(k) t}\right)-2 M_g(k) \propto X(k)+X^{\dagger}(k) 
    \\  \Leftrightarrow 
    \frac{d}{dt}\left(e^{X t} \widetilde{\Delta}(0) e^{X^{\dagger}t}\right) = (2-c)M_g(k_0)-cM_l^T(-k_0),
\end{eqnarray}
This can be shown as following    
\begin{eqnarray}
& \frac{d }{d t} \Delta(t)=X \Delta(t)+\Delta(t) X^{+}+2 M _g \\
\Leftrightarrow & \frac{d }{d t} \Delta(t)-2 M _g=X \Delta(t)+\Delta(t) X^{\dagger}
\end{eqnarray}
At dynamical topological phase transition\\
\begin{eqnarray}
&\Delta(t,k) \propto \mathbb{I} (k)
\end{eqnarray}
so
\begin{eqnarray}
&\frac{d }{d t} \Delta(t)-2 M _g \propto X+X^{\dagger}\,.
\end{eqnarray}
Combining with
\begin{eqnarray}
&\Delta(t)= e^{X t} \widetilde\Delta(0) e^{X^{\dagger}t} +\Delta_s\\
&\frac{d }{d t} \Delta s=0 
\end{eqnarray}
we get Eq. (\ref{dnpt}).

\bibliography{TopologyinLindbladequations}
\end{document}